\documentclass[preprint]{aastex}
\newcommand{\eg}{e.$.\!$g.\ } 
\newcommand{\be}{\begin{equation}}
\newcommand{\ee}{\end{equation}}
\newcommand{\etal}{ et al.\ }
\newcommand{\ie}{i$.\!$e$.\!$, }

\newcommand{\unit}[1]{\,{\rm #1}}
\newcommand{\hubble}[1]{H_0={#1}\unit{km\,sec^{-1}\,Mpc^{-1}}}

\newcommand{\labequn}[1]{\label{eq:#1}}

\newcommand{\labfig}[1]{\label{fig:#1}}
\newcommand{\labsecn}[1]{\label{sec:#1}}

\newcommand{\mean}[1]{\left\langle#1\right\rangle}

\newcommand{\mubzero}{\mu_{\rm b}(0)}
\newcommand{\re}{r_{\rm e}}
\newcommand{\idr}{I_{\rm d}(r)}
\newcommand{\idzero}{I_{\rm d}(0)}
\newcommand{\rd}{r_{\rm d}}
\newcommand{\rbyrdl}{r/\rd}
\newcommand{\exprbyrdl}{e^{-(\rbyrdl)}}
\newcommand{\mmubre}{\mean{\mubre}}
\newcommand{\mubre}{\mu_{\rm b}(\re)}
\newcommand{\const}{{\rm constant}}

\newcommand{\myequn}[2]{\begin{equation}{#2}
\labequn{#1}
\end{equation}}

\newcommand{\bea}{\begin{eqnarray}}
\newcommand{\eea}{\end{eqnarray}}

\begin{document}
\title{A NEAR INFRARED PHOTOMETRIC PLANE FOR ELLIPTICALS AND BULGES OF SPIRALS}
\author{Habib G. Khosroshahi \altaffilmark{1,2}, Yogesh Wadadekar
\altaffilmark{1}, Ajit Kembhavi \altaffilmark{1} and Bahram Mobasher\altaffilmark{3}}
\altaffiltext{1}{
Inter University Centre for Astronomy and Astrophysics, Post Bag
4, Ganeshkhind, Pune 411007, India; habib@iucaa.ernet.in, yogesh@iucaa.ernet.in, 
akk@iucaa.ernet.in}
\altaffiltext{2}{Institute for Advanced Studies in Basic Sciences,
P. O. Box 45195-159 Zanjan, Iran; khosro@iasbs.ac.ir}
\altaffiltext{3}{Astrophysics Group, Blackett Laboratory, Imperial College, 
Prince Consort Road, London, SW7 2BZ, UK; b.mobasher@ic.ac.uk}

\begin{abstract}

We report the existence of a single plane in the space of global
photometric parameters describing elliptical galaxies and the bulges of
early type spiral galaxies.  The three parameters which define the
plane are obtained by fitting the Sersic form to the brightness
distribution obtained from near-infrared K band images.  We find, from
the range covered by their shape parameters, that the elliptical galaxies form
a more homogeneous population than the bulges.  Known correlations
like the Kormendy relation are projections of the photometric plane.
The existence of the plane has interesting implications for bulge
formation models.
  
\end{abstract}


\section{INTRODUCTION}

Amongst the most important issues in studying formation of galaxies
are the epoch and physical mechanism of bulge formation.  There are
two competing scenarios for the formation of bulges. One assumes that
the bulge and disk form independently, with the bulge preceding the
disk (\eg Andredakis, Peletier \& Balcells 1995, hereafter APB95),
while the other suggests that the disk forms first and the bulge
emerges later from it by secular evolution (Courteau, de Jong \&
Broeils 1996).  However, recent analysis of a complete sample of early
type disk galaxies (Khosroshahi, Wadadekar \& Kembhavi 2000,
hereafter KWK) has shown that more than one mechanism of bulge
formation may be at work. This is corroborated by recent HST
observations, which show that distinct bulge formation mechanisms
operate for large and small bulges.

Recent studies have revealed that the bulges of {\em early} type disk
galaxies are old (Peletier \etal 1999).  This is in agreement with
semi-analytical results which claim that the bulges of field as well
as cluster disk galaxies are as old as giant elliptical galaxies in clusters
(Baugh \etal 1998).  However, the formation mechanism of the
bulges of {\em late} type spiral galaxies is likely to be very
different. For example, Carollo (1999) found that although a small
bulge may form at early epochs, it is later fed by gas flowing into
the galaxy core, possibly along a bar-like structure caused by
instabilities in the surrounding disk.  The situation becomes more
complicated for intermediate sized bulges, with some having formed at
early epochs and some relatively recently from gas inflows.

Correlations among global photometric parameters that characterize the
bulge -- such as colors, scale lengths etc. -- can be used to
differentiate between the competing bulge formation models. These have
the advantage of being independent of spectroscopic parameters such as
velocity dispersion which are difficult to measure for bulges.  Some
of the photometric parameters such as the colors are measured
directly, while others like the scale lengths require elaborate
bulge-disk decomposition using empirical models for the bulge and disk
profiles.  Conventionally, radial profiles of bulges (de Vaucouleurs
1959), like those of elliptical galaxies (de Vaucouleurs 1948), have
been modeled by the $r^{1/4}$ law.  These are fully described by two
parameters determined from the best fit model -- the central surface
brightness $\mubzero$ and an effective radius $\re$, within which half
the total light of the galaxy is contained. In recent years an
additional parameter has been introduced, with the $r^{1/4}$ law
replaced by a $r^{1/n}$ law (Sersic 1968), where $n$ is a free
parameter (\eg Caon \etal 1993, APB95, KWK). The so called Sersic
shape parameter $n$ is well correlated with other observables like
luminosity, effective radius, the bulge-to-disk luminosity ratio and
morphological type (APB95). In particular, it has been demonstrated that a
tight correlation exists between $\log n$, $\log \re$ and $\mubzero$ (KWK).
 
In this letter we show that $\log n$, $\log \re$ and $\mubzero$ for
elliptical galaxies are tightly distributed about a plane in
logarithmic space, and that this {\it photometric plane} for
elliptical galaxies is indistinguishable from the analogous plane for the
bulges of early type disk galaxies.  Throughout this paper we use
$\hubble{50}$ and $q_0=0.5$.

\section{THE DATA AND DECOMPOSITION METHOD}
\labsecn{data}

The analysis in this study is based on the near-IR, K-band images of
42 elliptical galaxies, in the Coma cluster (Mobasher \etal 1999).
This is combined with a complete magnitude and diameter limited sample
of 26 early type disk galaxies in the field (Peletier and Balcells
1997) from the Uppsala General Catalogue (Nilson 1973). Details about
the sample selection, observations and data reduction are given in the
above references.  We chose to work with K band images because the
relative lack of absorption related features in the band leads to
smooth, featureless light profiles which are convenient for extraction
of global parameters.

Extracting the global bulge parameters of a galaxy requires the
separation of the observed light distribution into bulge and disk
components.  This is best done using a 2-dimensional technique, which
performs a $\chi^2$ fit of the light profile model to the galaxy
image.  A scheme is used in which each pixel is weighted by its
estimated signal-to-noise ratio (Wadadekar, Robbason and Kembhavi
1999).
 
We decomposed all the galaxies in our sample into a bulge component
which follows a $r^{1/n}$ law with
\myequn{ibulge}
{I_{bulge}(r) = I_b(0) e^{ -2.303 b_n (r/\re)^{1/n}},} where $b_n =
0.8682 n - 0.1405$, and $r$ is the distance from the center along the
major-axis.  The disk profile is taken to be an exponential
$\idr=\idzero\exprbyrdl$, where $\rd$ is the disk scale length and
$\idzero$ is the disk central intensity.  Apart from the five
parameters mentioned here, the fit also involves the bulge and disk
ellipticities. The model for each galaxy was convolved with the
appropriate  point spread function (PSF).  Details of the procedure used 
in the decomposition of
the disk galaxies are given in KWK.  We have got good fits for all 42
of the 48 elliptical galaxies in the complete sample of Mobasher \etal
(1999) for which we have data, and for 26 of the 30 disk galaxies in
the complete sample of APB95. Twelve of the 42 elliptical galaxies show a
significant disk ($ D/B \gtrsim 0.2 $). Four disk galaxies did not
provide good fits because of their complex morphology (see KWK) and
these galaxies have been excluded from our discussion.  

In Figure 1 we plot histograms showing the distribution of the shape
parameter $n$ for the two samples. For elliptical galaxies, $n$ ranges from
1.7 to 4.7 with a clear peak around $n=4$. This observation
is in agreement with the fact that de Vaucouleurs' law has
historically provided a reasonable fit to the radial profile of most (but
not all) elliptical galaxies.  For the disk galaxies $n$ ranges from
1.4 to 5 with an almost uniform distribution within this range. The
rather flat distribution in $n$ for the disk galaxies implies that 
de Vaucouleurs' law will provide a poor fit to the bulges of these
galaxies. This is indeed the case as demonstrated in de Jong (1996).

\section{THE PHOTOMETRIC PLANE} 
\labsecn{correlations}

Study of the correlations among the parameters describing photometric
properties of elliptical galaxies and bulges of spiral galaxies is
essential in constraining galaxy formation scenarios.  We find that
for the spiral galaxies sample, the bulge central surface brightness
is well correlated with $\log n$, with a linear correlation
coefficient of -0.88 (Figure 2).  The corresponding coefficient for
the elliptical galaxies, also shown in Figure 2, is -0.79. These
relations are significant at $>99.99$\% level as measured by the
Student's $t$ test.  There is a weak correlation between bulge
effective radius and $n$ for the disk galaxies (KWK) but such a
correlation does not exist for the elliptical galaxies.

An anti-correlation between the effective radius and mean surface
brightness within the effective radius -- known as the Kormendy
relation (Kormendy 1977, Djorgovski \& Davis 1987)-- has been reported
in elliptical galaxies. In Figure 3 we plot the mean surface
brightness within the effective radius against effective radius for
the two samples.  The elliptical galaxies are clustered around the
best fit line -- $\mmubre = 2.57 \log\re + 14.07$ with rms scatter of
0.59 in mean surface brightness.  A weaker relation with larger
scatter exists for the bulges of the early type spiral galaxies
suggesting a formation history similar to that of elliptical galaxies.
As we demonstrated in KWK, the bulges of late type spiral galaxies do
{\em not} show a Kormendy type relation, suggesting a different
formation history.

It is possible that some of the scatter seen in the Kormendy
relation is caused by the effect of a third parameter, which can only
be $n$ in our scheme.  We have applied standard bivariate analysis
techniques to obtain the best fit plane in the space of the three
parameters $\log n$, $\mubzero$ and $\log\re$.  We find that the least
scatter around the best fit plane is obtained by expressing it in the
form $\log n= A\log\re + B\mubzero + \const$, and minimizing the
dispersion of $\log n$ as measured by a least-squares fit.

The equation of 
the best fit plane for the elliptical galaxies is  
\bea
\log n &=& (0.173 \pm 0.025)\log \re - (0.069 \pm 0.007) \mubzero + (1.18 \pm 
0.05),
\eea
while for the bulges of the disk galaxies it is 
\bea
\log n &=& (0.130 \pm 0.040)\log \re - (0.073 \pm 0.011) \mubzero +
(1.21 \pm 0.11).
\eea
The errors in the best fit coefficients here were obtained by fitting  planes
to synthetic data sets generated using the bootstrap method with
random replacement (Fisher 1993).  The scatter in $\log n$ for the above
planes is 0.043 dex (corresponding to 0.108 magnitude) and 0.058
dex (corresponding to 0.145 magnitude)  respectively.
 
The angle between the two planes is $2.41 
\pm 1.99\deg$; this error was also obtained by the bootstrap
technique. The difference in angle between the two planes is only slightly more
than the $1\sigma$ uncertainty, which strongly suggests that the two
planes are identical. We therefore obtained a new equation for 
the common plane, combining the data for the two samples, which is:
\bea
\log n &=& (0.172 \pm 0.020)\log \re - (0.069 \pm 0.004) \mubzero + (1.18 \pm 
0.04),
\eea  
The smaller errors here are due to the increased size of the combined
sample.  

A face-on and two mutually orthogonal edge-on views of the best fit
plane for the two samples are shown in Figure 4. ${\rm K_1}, {\rm K_2}$
and ${\rm K_3}$ are orthonormal vectors constructed from linear
combinations of the parameters of the photometric plane.  An
additional representation of this best fit plane with $\log n$ as the ordinate,
together with the data points used in the fit, is shown in Figure 5.
The rms scatter in $\log n$ here is 0.050 dex, corresponding to 0.125
magnitude.  This is comparable to the rms error in the fitted values
of $\log n$, so any intrinsic scatter about the plane is small.

It is possible that some of the observed correlation is produced due
to correlations between the fitted parameters of the bulge-disk
decomposition.  We have examined the extent of such an induced
correlation, using extensive simulations of model galaxies obtained
using the observed distributions of $n$, $\mubzero$ and $\re$ for both
samples. We chose at random a large number of $n$, $\mubzero$ and
$\re$ triplets from these distributions, with the values in each
triplet chosen independently of each other.  Such a random selection
ensured that there was no correlation between the input
parameters. Other parameters needed to simulate a galaxy, like disk
parameters, were also chosen at random from the range of observed
values.  We added noise at the appropriate level to the simulated
images and convolved the models with a representative point spread
function.  We then extracted the parameters for these galaxies using
the same procedure as we adopted for the observed sample.  Results
from the fit to the simulated data do not show significant univariate
or bivariate correlations between the extracted parameters.  This
indicates that the correlations seen in the real data are not
generated by correlated errors.

\section{DISCUSSION}

It is tempting to investigate the use of Equation 4 as a distance
indicator.  The main source of uncertainty here is that the two
distance independent parameters, $\log n$ and $\mubzero$, are in fact
correlated, leading to an increased error in the best-fit photometric
plane, and hence in the estimated $\log\re$ values.  This gives an
error of 53\% in the derived distance, which is similar to the error
in other purely photometric distance indicators, but is significantly
larger than the $\sim20\%$ error found in distances from the the
near-IR fundamental plane, using both photometric and spectroscopic
data (\eg Mobasher \etal 1999; Pahre, Djorgovski \& de Carvalho
1998). However, data for the photometric plane are easy to obtain, as
no spectroscopy is involved and it should be possible to get more
accurate distances to clusters by independently measuring distances to
several galaxies in the cluster.

The elliptical galaxies seem to form a more homogeneous population
than the bulges of spiral galaxies, as revealed from the distribution
of their shape parameter, $n$.  Considering that the near-infrared
light measures contribution from the old stellar population in galaxies
(\ie the integrated star formation) and since the near-infrared mass
to luminosity ratio $(M/L)_K$ is expected to be constant among the
galaxies, the relatively broad range covered by the shape parameter,
$n$, for bulges of spirals reveals differences in the distribution of
the old population among these bulges. Environmental factors
could play an important role here, since the elliptical galaxies in
our sample are members of the rich Coma cluster while the spiral
galaxies are either in the field or are members of small groups. While
elliptical galaxies and bulges appear to be different in the context
of the Kormendy relation, they are unified onto a single plane when
allowing for differences in their light distribution, as measured by
the shape parameter, $n$.  This supports the use of $n$ as a
fundamental parameter in studying elliptical galaxies and bulges of
early type disk galaxies, similar to the velocity dispersion in the
fundamental plane of elliptical galaxies. The existence of a
photometric plane for ellipticals and bulges of early-type disk
galaxies further supports an independent study by Peletier \etal
(1999) which found that bulges in early type disk galaxies and
ellipticals have similar stellar content and formation epochs. It will
be important to see whether the photometric plane for lenticulars too
coincides with the plane for elliptical galaxies and bulges of early
type galaxies, to explore whether lenticulars indeed provide an
evolutionary link between elliptical galaxies and early type disk
galaxies.
 
The observed tightness of the photometric plane provides a strong
constraint on formation scenarios, and therefore it is required to
study its physical basis.  Recently Lima Neto, Gerbal \& Marquez
(1999) have proposed that elliptical galaxies are stellar systems in a
stage of quasi-equilibrium, which may in principle, have a unique
entropy per unit mass -- the {\em specific entropy}. It is possible to
compute the specific entropy assuming that elliptical galaxies behave
as spherical, isotropic, one-component systems in hydrostatic
equilibrium, obeying the ideal-gas equations of state. Using the
specific entropy and a analytic approximation to the three dimensional
deprojection of the Sersic profile, they predict a relation between
the three parameters of the Sersic law. This relation defines a plane
in parameter space which they call the {\em entropic plane}. The
parameters used in their fit are not identical to ours, and therefore
a comparison is not straightforward. 

The photometric plane may be useful in probing the bulge
formation mechanism in galaxies.  In this context it will be interesting to
see whether the bulges of late type disk galaxies also share a single
plane with the bulges of early type disk galaxies and ellipticals.
If they do, then a single mechanism
for bulge formation in all types would be indicated.  But if bulges in
early and late type disk galaxies are formed differently (Peletier \etal 1999,
Carollo 1999) then a single plane is not expected.
It will also be of interest to compare scaling laws which follow from
the photometric plane with those implied by the existence of the
fundamental plane (Djorgovski \& Davis 1987) of elliptical galaxies.
 
\acknowledgements
We thank S. George Djorgovski for useful discussions and
Y. C. Andredakis, R. F. Peletier and M. Balcells for making their data
publicly available. We thank an anonymous referee for comments that
helped improve this paper. One of us, HGK, would like to thank Y. Sobouti and
J. V. Narlikar for their help and support during this project.
\newpage
\centerline{REFERENCES}
\noindent 
Andredakis, Y.C., Peletier, R.F., \& Balcells, M. 1995, \mnras, 275,
874 (APB95)\\
Baugh, C.M., Cole, S., Frenk, C.S. \& Lacey, C.G. 1998, \apj, 498, 504\\ 
Caon, N., Capaccioli, M. \& D'Onofrio, M. 1993, \mnras, 163, 1013\\
Carollo, C. M. 1999, \apj, 523, 566\\ 
Courteau, S., de Jong, R. S. \& Broeils, A. H. 1996, \apj, 457, L73\\
de Jong R. S. 1996, \aaps, 118, 557 \\
de Vaucouleurs, G. 1948, Ann. d'Astrophys., 11, 247 \\
de Vaucouleurs, G. 1959, Hdb. d. Physik, 53, 311 \\
Djorgovski, S.G. \& Davis, M. 1987, \apj, 313, 59\\
Fisher, N.I. 1993, Statistical analysis of circular data, 
(Cambridge: Cambridge University Press)\\
Khosroshahi H. G., Wadadekar, Y. \& Kembhavi A. 2000, \apj, in press 
(astro-ph/9911402) (KWK)\\
Kormendy, J. 1977, \apj, 217, 406 \\
Lima Neto, G. B., Gerbal, D., \& Marquez, I. 1999, \mnras, 309, 481\\
Mobasher, B., Guzman, R., Aragon-Salamanca, A., \& Zepf, S. 1999,
\mnras, 304, 225\\
Nilson, P. 1973, Uppsala General Catalogue of Galaxies, (Uppsala: Astronomiska 
Observatorium)\\
Pahre, M. A., Djorgovski, S. G., \& de Carvalho, R. R. 1998, \aj, 116, 1591\\
Peletier, R.F., \& Balcells, M. 1997, New Astronomy, 1, 349 \\
Peletier, R.F., Balcells, M., Davies, R.L., Andredakis, Y. 1999, \mnras, 
in press (astro-ph/9910153)\\
Sersic, J.L. 1968, Atlas de galaxies australes. Observatorio Astronomica, 
Cordoba\\
Wadadekar Y., Robbason R., \& Kembhavi, A. 1999, \aj, 117, 1219\\
\newpage
\begin{figure}
\plotone{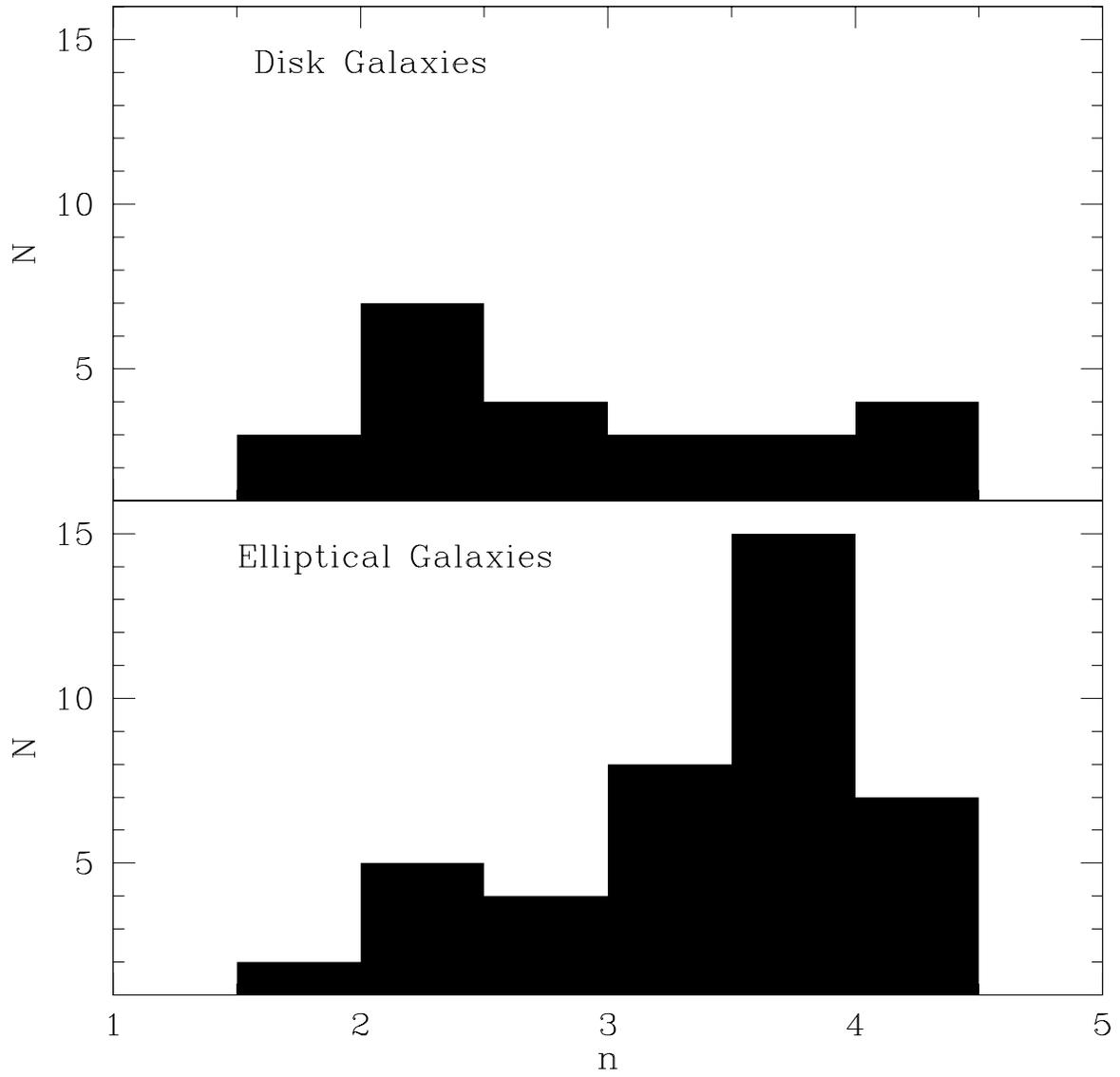}
\caption{Histogram of the shape parameter $n$ for bulges of disk galaxies and 
elliptical galaxies.}
\labfig{histogram}
\end{figure}

\newpage
\begin{figure}
\plotone{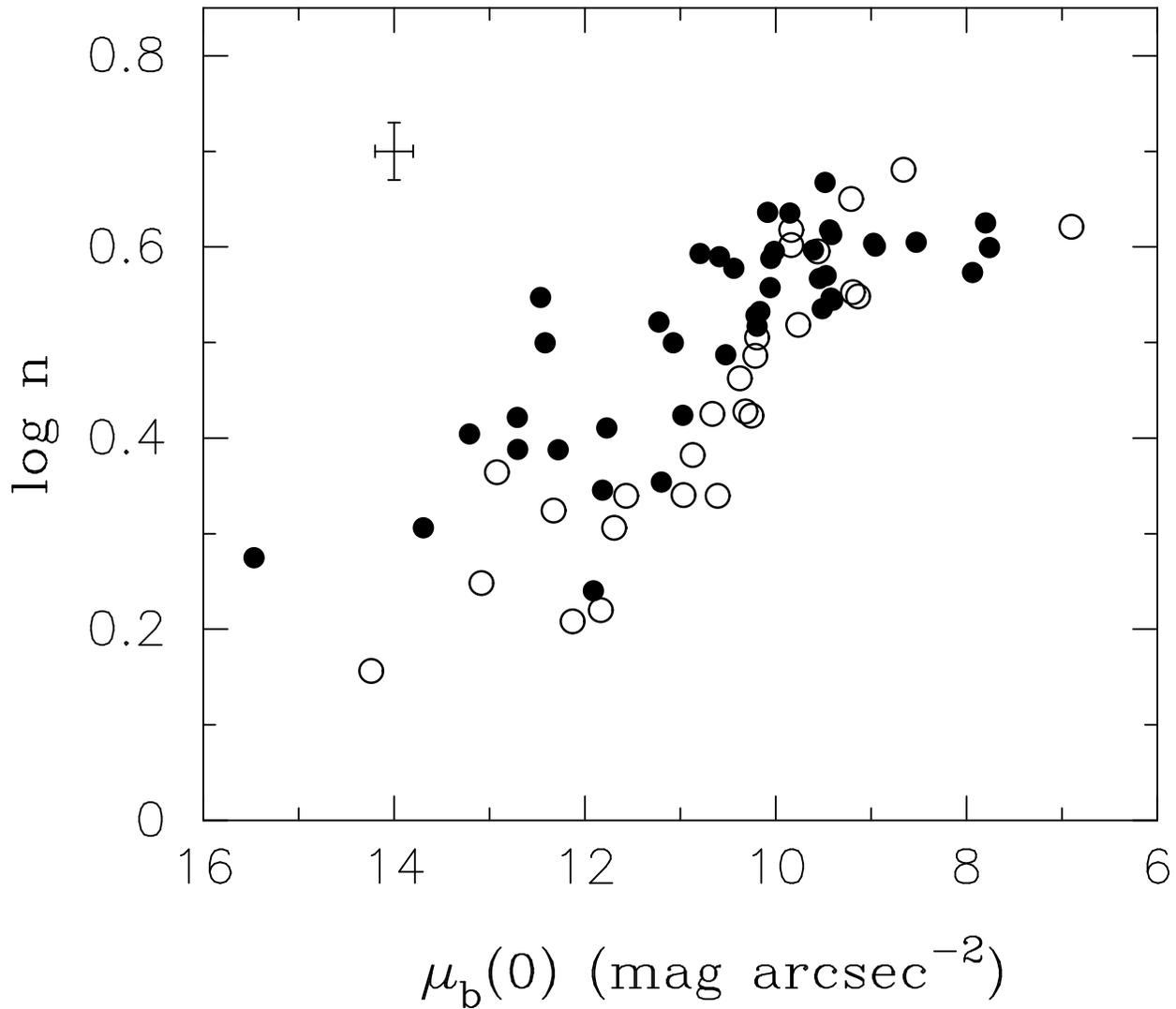}
\caption{Logarithm of the shape parameter, $n$, is plotted against the 
deconvolved bulge central surface brightness.
The open circles represent bulges of disk galaxies and the filled circles
represent elliptical galaxies. Typical error bars are shown.}
\labfig{nmu0}
\end{figure}

\newpage
\begin{figure}
\plotone{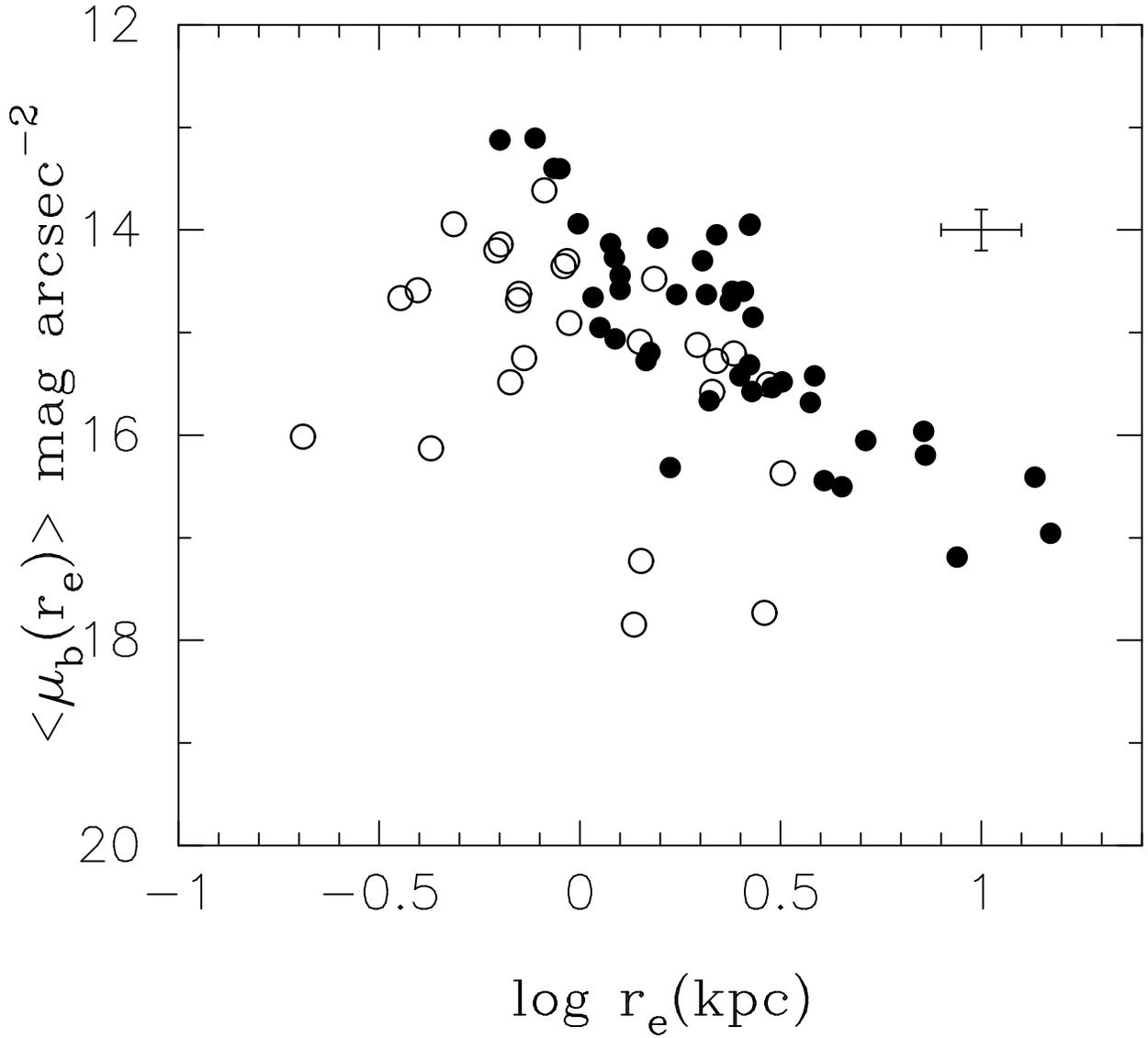}
\caption{Kormendy relation for bulges of disk galaxies and
elliptical galaxies. The filled circles represent elliptical galaxies and the open
circles represent bulges of disk galaxies. Typical error bars are shown.}
\labfig{kormendy}
\end{figure}

\newpage
\begin{figure}
\plotone{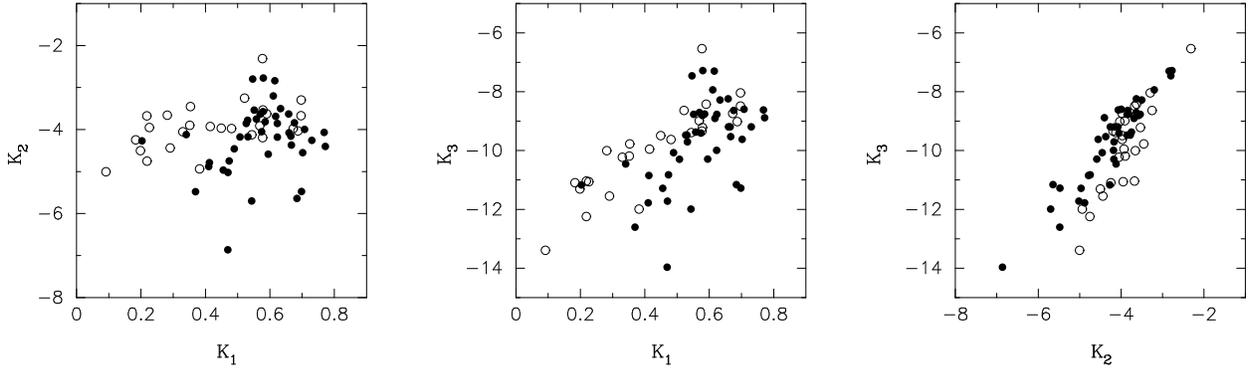}
\caption{ A face-on and two edge-on views of the {\em photometric plane}, where 
${\rm K_1} = 0.986 \log n + 0.169 \log r_e$, 
${\rm K_2} = 0.157 \log n - 0.916 \log r_e - 0.377 \mubzero$ and
${\rm K_3} = -0.064 \log n + 0.372 \log r_e - 0.93 \mubzero$. 
The filled circles represent elliptical galaxies and the open
circles represent bulges of disk galaxies.}
\labfig{3dplane}
\end{figure}

\newpage
\begin{figure}
\plotone{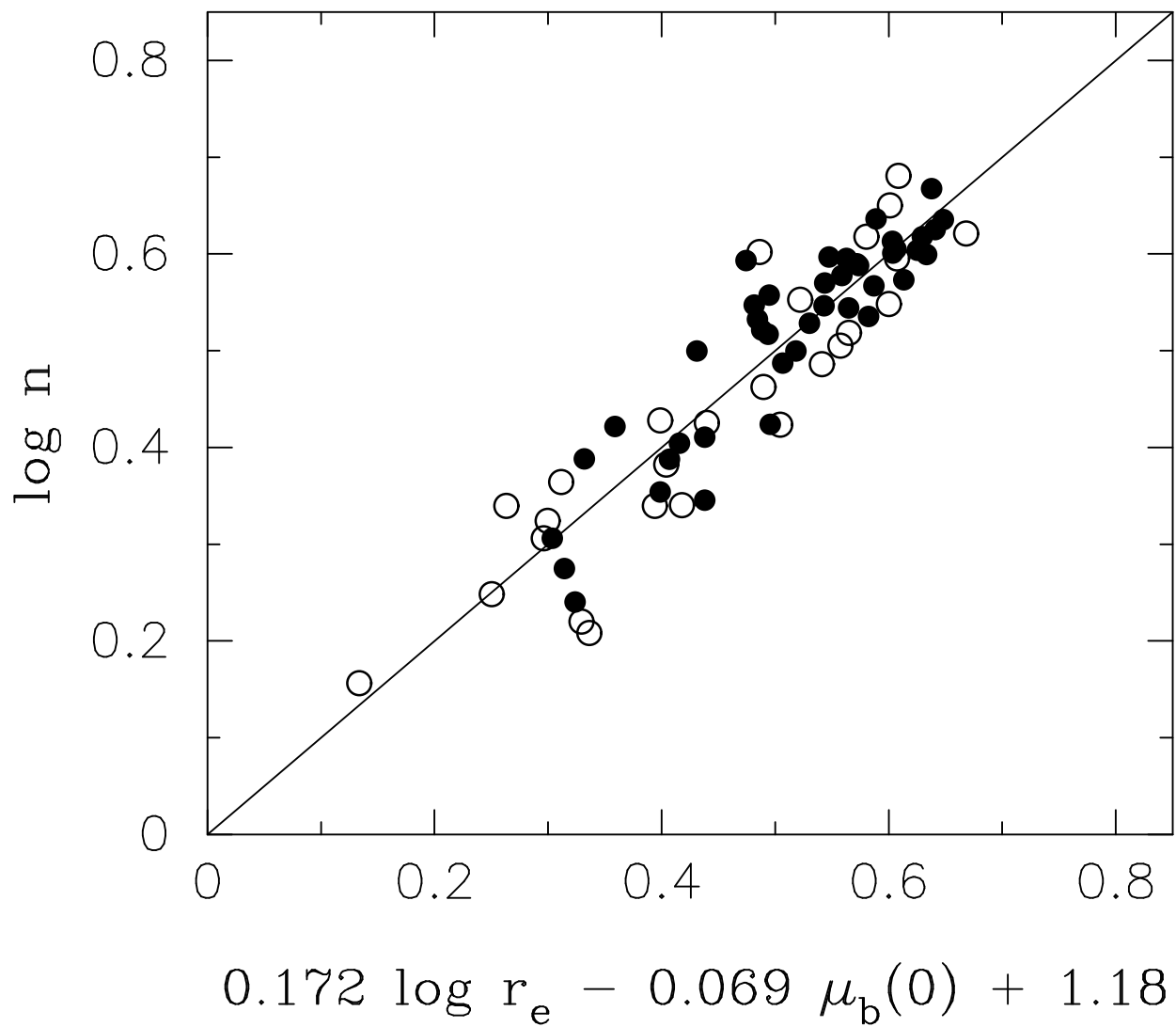}
\caption{A tight representation of the best fit {\em photometric plane}.  
The filled circles represent elliptical galaxies and the open circles 
represent bulges of disk galaxies. The line is a plane fit to the data.}
\labfig{bicor}
\end{figure}

\end{document}